\documentclass{elsart}

\usepackage{natbib}

\usepackage{graphicx}
\usepackage{epsfig}

\usepackage{amssymb}

\begin{document}

\def\lsim{\mathrel{\lower .85ex\hbox{\rlap{$\sim$}\raise
.95ex\hbox{$<$} }}}
\def\gsim{\mathrel{\lower .80ex\hbox{\rlap{$\sim$}\raise
.90ex\hbox{$>$} }}}

\begin{frontmatter}



\title{Black Hole Spectral States and Physical Connections}


\author{John A. Tomsick}

\address{Center for Astrophysics and Space Sciences, Code
0424, University of California at San Diego, La Jolla, CA,
92093, USA (e-mail: jtomsick@ucsd.edu)}

\begin{abstract}

The dramatic changes seen in the X-ray spectral and timing
properties of accreting black hole candidates (BHCs) provide 
important clues about the accretion and jet formation processes 
that occur in these systems.  Dividing the different source 
behaviors into spectral states provides a framework for studying 
BHCs.  To date, there have been three main classification
schemes with Luminosity-based, Component-based, or 
Transition-based criteria.  The canonical, Luminosity-based 
criteria and physical models that are based on this concept do 
not provide clear explanations for several phenomena, including 
hysteresis of spectral states and the presence of jets.  I 
discuss the re-definitions of states, focusing on an application 
of the Component-based states to more than 400 {\em RXTE} 
observations of the recurrent BHC 4U~1630--47.  We compare the 
X-ray properties for the recent 2002--2004 outburst to those of 
an earlier (1998) outburst, during which radio jets were observed.
The results suggest a connection between hysteresis of states 
and major jet ejections, and it is possible that both of these 
are related to the evolution of the inner radius of the optically 
thick accretion disk.

\end{abstract}


\end{frontmatter}

\section{Introduction}

It has long been known that accreting black hole candidates (BHCs)
exhibit a wide array of emission properties, and, over the years, 
these properties have been divided in various ways into spectral 
states.  The first state transition was detected in the early 1970s
using {\em Uhuru} when the 2--6 keV flux from Cyg X-1 was seen to
decrease by a factor of four while the 10--20 keV flux increased
by a factor of two.  Upon the discovery of this change, 
\cite{tananbaum72} called this a transition from a ``low'' state 
to a ``high'' state based on the change in the 2--6 keV flux.  
It was also realized at this time that the radio properties can 
be associated with the X-ray states as \cite{hw71} also reported 
a radio detection from Cyg X-1 that \cite{tananbaum72} associated 
with the low state.  Since these early discoveries, more sensitive, 
higher-throughput, broadband X-ray detectors and multi-wavelength 
coverage have provided a much more detailed understanding of the 
phenomenology of spectral states, including both the spectral 
\citep[e.g.,][]{wr83,mitsuda84,ts96,grove98} and timing 
\citep[e.g.,][]{miyamoto91,vdk95} properties.  Also, theoretical 
work has provided insights into the physical processes that are 
occurring in these binary accreting BHC systems 
\citep[e.g.,][]{ss73,abramowicz88,emn97}.  

Some of the best examples of broadband, 1--2000 keV, energy 
spectra combine spectra from the {\em Compton Gamma-Ray
Observatory's (CGRO)} OSSE instrument with spectra from 
various soft X-ray instruments \citep{grove98}.  From these, 
it is clear that there are three components that dominate BHC 
spectra:  The soft component; the cutoff power-law component; 
and the steep power-law component.  The soft component is due 
to thermal emission from an optically thick accretion disk.
In many cases, the shape of this component is consistent with 
the predictions for a \cite{ss73} disk with an inner disk
temperature of $\sim$1~keV.  Historically, the cutoff 
power-law component has been explained by invoking inverse
Componization from a hot electron ``corona'' with a thermal 
energy distribution at a temperature of $\sim$100~keV.  
However, the geometry of the corona is not known, and the
question of whether jets may be involved in X-ray production 
is a current topic of debate \citep{mn04}.  The origin of 
the steep power-law component is also a topic of debate
\citep[see, e.g.,][]{mr03}.  Although the emission mechanism 
may be inverse Comptonization, the fact that the component
extends to MeV energies or beyond without a cutoff 
\citep{tomsick99} makes it unlikely that it comes from a 
thermal electron distribution.  Hybrid thermal/non-thermal 
models have been developed \citep{coppi92}, but the mechanism 
for accelerating the electrons is still unclear.

When the thermal component is present, it is usually 
accompanied by the steep power-law component at some level, 
and if the thermal component dominates the 2--10 keV band, 
it is traditionally said that the source is in the 
``High-Soft'' state.  At other times, when the cutoff
power-law component dominates the entire X-ray spectrum,
this state has come to be known as the ``Low-Hard'' state.
The high throughput X-ray instruments such as {\em Ginga}
and the {\em Rossi X-ray Timing Explorer (RXTE)} have made
it clear that the X-ray timing properties are closely 
related to the spectral properties, and they are also
important in defining the spectral states.  While strong
timing noise is seen in the High-Soft state and weak noise
is seen in the Low-Hard state, it was noticed by 
\cite{miyamoto91} and others that at very high luminosities, 
the power spectra typically show intermediate levels of
timing noise and usually exhibit quasi-periodic oscillations 
(QPOs).  This state became known as the Very High State
(VHS).  In addition to the VHS, the presence of another
distinct state, the ``Intermediate'' state (IS), with 
properties (luminosity, timing noise, spectral hardness) 
intermediate to the Low-Hard and High-Soft states was 
suggested \citep{belloni96,mendez97}.  When the very low 
luminosity ``Off'' or ``Quiescent'' state is included, 
there are five canonical spectral states:  VHS, High-Soft, 
IS, Low-Hard, and Quiescent.

Based on this Luminosity-based classification of states, 
a physical picture including advection-dominated accretion 
flows (ADAFs) was advanced to explain many of the X-ray
properties \citep{emn97}.  This picture supposes that all
of the emission properties are set by changes in the mass
accretion rate ($\dot{M}$).  While other physical parameters, 
such as the inner radius of the optically thick disk and
the size of the corona might change, the concept of the 
model is these are tied to $\dot{M}$.  While relatively 
successful in describing some of the spectral properties of 
BHCs, this model clearly misses some of the important BHC 
physics.  In the following, I describe some of the evidence 
for other physics.  Then, I discuss state classifications 
that have been recently advanced in an attempt to adapt the 
spectral state definitions in light of recent observational 
results.  Finally, I use a large set of {\em RXTE} 
observations of the BHC transient 4U~1630--47 as a case 
study and discuss the results.

\section{Beyond the Canonical Spectral States}

\subsection{Radio Connections to States}

Although it has been known for some time that many of the 
BHC systems are radio emitters, most of the systematic 
studies of how the radio properties are connected to the
spectral states have only been carried out over the past
several years.  It has now been well-established that 
steady, flat-spectrum radio emission is associated with 
the Low-Hard state \citep{corbel00,fender01}.  For two
sources, this radio emission is resolved, indicating that
radio jets are present \citep{stirling01,dhawan00}.
In many cases, the radio and X-ray fluxes are strongly 
correlated \citep{gfp03}, indicating that the radio and
X-ray emission mechanisms are closely linked.  The 
presence of a steady jet in the Low-Hard state has 
implications for the \cite{emn97} physical model as this 
state is where the ADAF forms when accretion energy is 
advected into the black hole.  It has been suggested that 
this accretion energy may be lost to the jet instead 
\citep{bb99,fgj03}.

\begin{figure}
\centerline{\includegraphics[width=4.5in,angle=0]{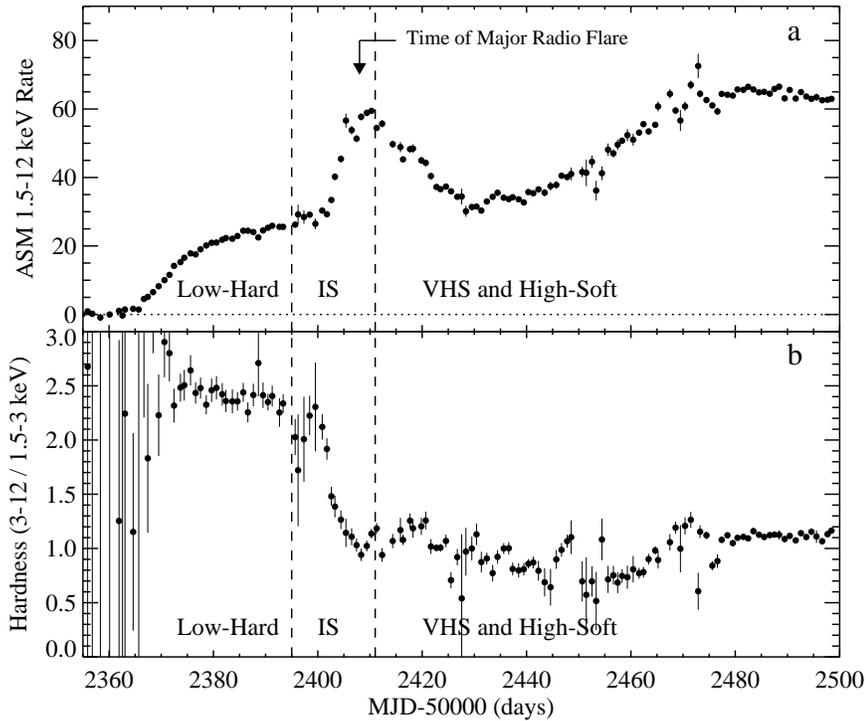}}
\caption{Demonstration of X-ray and radio evolution that
has been seen for several BHCs.  The light curve ({\em a}) 
and hardness ratio values ({\em b}) are for the BHC GX~339--4.  
The figure has been adapted from \cite{corbel04}, and the 
time of the major radio ejection (marked with an arrow) 
comes from \cite{gallo04}.  For several systems, it
has been found that major radio ejections follow a bright
``Low-Hard'' state and occur close to a state transition.
\label{fig:gx339}}
\end{figure}

In addition to the steady radio jets, BHC systems show
major, discrete jet ejections when the systems are at
high X-ray fluxes (likely at high $\dot{M}$).  
Although these major jet ejections were initially given
more attention than the steady jets likely due to 
apparently superluminal motions measured for some of the 
discrete ejections \citep{mr99}, a clear connection 
between the ejections and specific spectral states has 
been lacking.  With studies of several recent BHC 
outbursts where major ejections occurred, this connection 
may be becoming clearer.  The example in 
Figure~\ref{fig:gx339} shows the evolution for the BHC 
GX~339--4 where the source started off in the Low-Hard 
state, made a transition to the IS, and then a major radio 
flare occurred when the source made a transition to the 
VHS, and similar evolution has been seen for several
other sources \citep{corbel04,fbg04}.

\subsection{Hysteresis and Possible Second Physical Parameters}

BHC transients show hysteresis in the evolution of their spectral
states during their rise and decay 
\citep[see][and references therein]{vdk04}.  Several examples
of hysteresis and luminosity-independence of states are given
in \cite{tomsick04_rossi}.  The most common form of hysteresis is
that the ``Low-Hard'' state sometimes occurs at the peak X-ray
luminosity for a given BHC outburst.  This hysteresis is not 
explained by physical models of BHCs such as the \cite{emn97} 
model described above, but it very likely indicates that there 
is at least one physical parameter that is at least partially 
independent of $\dot{M}$ that is important for setting the 
spectral state \citep{homan00,tomsick04_rossi}.  Several second parameters
are worth considering, including the presence and power of an 
outflow or jet, the fraction of accretion energy dissipated in 
the corona (vs.~the optically thick portion of the disk), the
accretion rate in the corona vs.~the accretion rate in the 
disk \citep[in two flow models such as][]{ct95}, and the inner
radius of the optically thick disk ($R_{\rm in}$), even though 
one might expect $R_{\rm in}$ to be at least partially dependent
on $\dot{M}$.

\section{Re-Definitions of States}

In light our changing view of spectral states and the physics
behind accreting BHCs, two re-definitions of BHC states have 
been advanced.  One of these re-definitions by \cite{mr03} is 
based on the details of the X-ray spectral and timing parameters 
and relies most heavily on the strength of the three spectral 
components described above.  Generally, sources are in the 
``Thermal-Dominant'' (TD) state when the soft component dominates;
the ``Steep Power-Law'' (SPL) state when the spectra have a 
strong power-law component with a photon index ($\Gamma$) steeper 
than 2.4; and the ``Hard'' state when the spectra are dominated
by a cutoff power-law component with $1.5 < \Gamma < 2.1$ and
the level of timing noise is high.  An important feature of the
\cite{mr03} classification scheme is that the authors have 
determined quantitative state definitions using 30 {\em RXTE}
spectra from 15 different BHC sources.  This gives precise 
meaning to the states and allows for clearer source-to-source
comparisons.  

Another re-definition scheme \citep{hb04,belloni05} groups the 
states by location in the hardness-intensity diagram according 
to where sharp changes in emission properties occur.  The 
following contains some discussion of this scheme, but we refer
to the paper in these proceedings by T.~Belloni for details.
In summary, three main classification schemes have been 
advanced:  The canonical definitions are {\em Luminosity-based};
the \cite{mr03} definitions are {\em Component-based}; and the
\cite{hb04} definitions are {\em Transition-based}.

\section{Application to 4U 1630--47}

The recurrent BHC transient 4U 1630--47 recently finished one
of its brightest and longest outbursts at the end of 2004.  It is
one of the most active BHC transients with 17 detected outbursts
going back to 1969.  The possibility to compare different outbursts
makes it an interesting source to study, and we have analyzed data
from over 400 {\em RXTE} observations taken during its 1998 
outburst \citep{tk00,dieters00,tbp01} and its recent 
outburst \citep{tomsick05}.  This provides an especially interesting 
comparison because strong and highly polarized radio emission was 
detected during the 1998 outburst (likely indicating the presence 
of radio jets) but not during its 2002--2004 outburst
\citep{hjellming99,hannikainen02,tomsick05}.  

\begin{figure}
\centerline{\includegraphics[width=5.0in,angle=0]{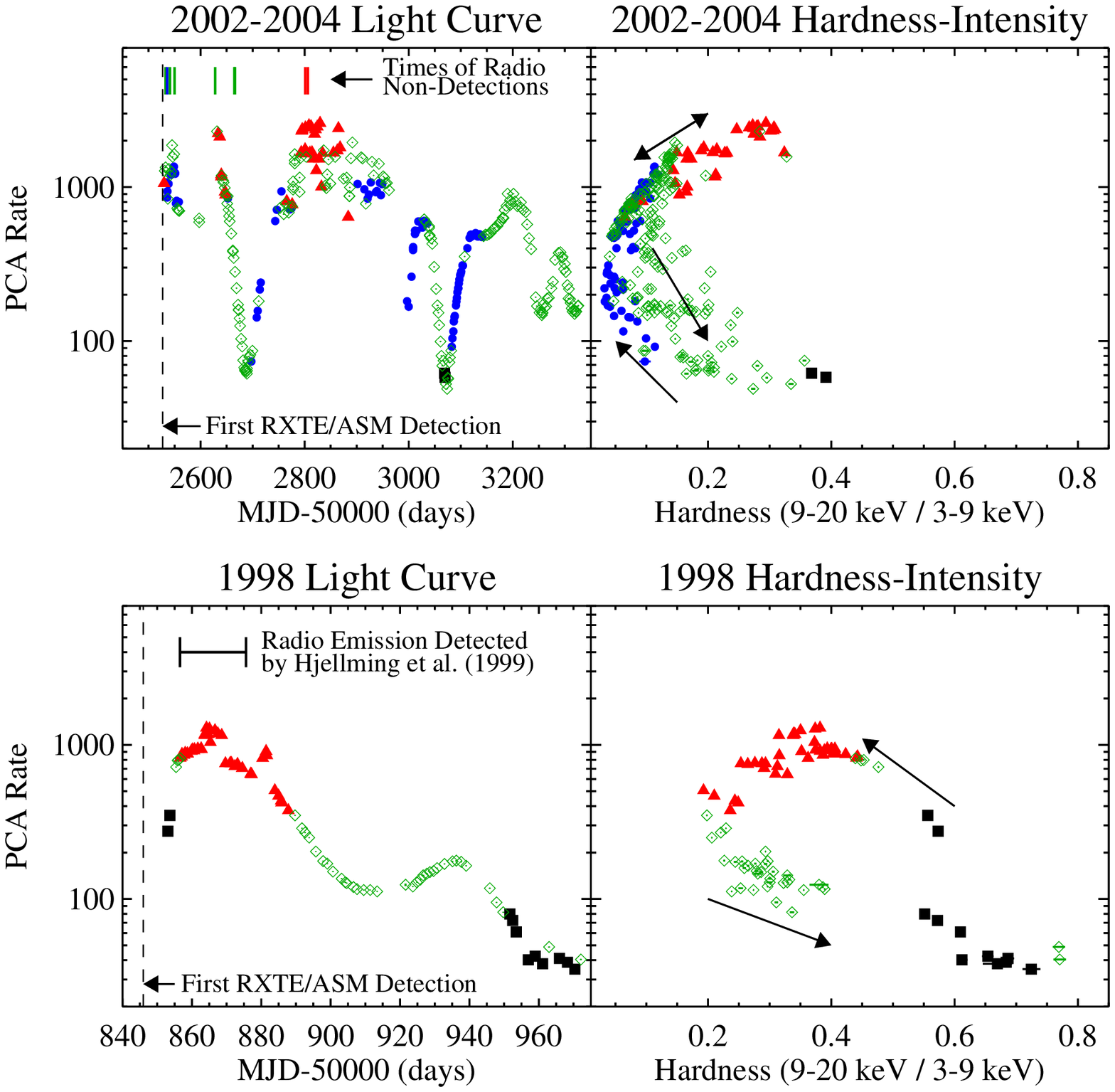}}
\caption{Properties of the 4U 1630--47 outbursts in 2002--2004
(top two panels) and 1998 (bottom two panels).  The data are
from pointed {\em RXTE} observations 
\citep[see][for details]{tomsick05}.  The colors/symbols 
correspond to the Component-based \cite{mr03} states as follows:  
Red triangles = Steep Power-Law; Green open diamonds = Intermediate 
State; Blue circles = Thermal-Dominant; Black squares = Hard state.  
For each outburst, 3--20 keV PCA light curves are shown along with
an indication of the results of radio observations, and
hardness-intensity diagrams are shown, with arrows indicating
the temporal evolution of the source in the diagram.\label{fig:4panel}}
\end{figure}

As described in detail in \cite{tomsick05}, we applied the
quantitative \cite{mr03} state definitions to 4U 1630--47
{\em RXTE} data.  The results of our comparison between the
2002--2004 and 1998 outbursts are shown in Figure~\ref{fig:4panel}.
The 1998 outburst follows a pattern that is now recognized
as being rather typical for BHCs (see Figure~\ref{fig:gx339}).
The source begins the outburst by reaching a relatively bright
Hard state before making a transition to an intermediate state
and then to a SPL state.  Radio emission likely signals a major
jet ejection starting around the time that the source makes a
transition to the SPL state.  The motion in the hardness-intensity
diagram is counter-clockwise, which is similar to the outbursts
that have been used as the basis for the Transition-based
state definitions \citep{hb04}.  The hysteresis in the Hard
state transition levels is clear with the transition from the
Hard state at the beginning of the outburst being a factor of
five or more higher in flux than the transition to the Hard state
at the end of the outburst.

The behavior during the 2002--2004 outburst is in sharp contrast
to the 1998 outburst.  The rise of the outburst is very fast, 
and there is no evidence that the source ever entered a bright
Hard state.  The overall outburst is considerably softer, with 
the source entering the TD state on several occasions.  Radio 
observations yielded non-detections for the TD, IS, and SPL states.  
The hardness-intensity diagram shows an evolution that is completely 
different from 1998, with the source being in the TD state during 
its rise.  The differences between outbursts indicates a connection 
between the X-ray behaviors and radio jets.  This connection may be 
related to the hardness of the outburst or to more subtle X-ray 
features such as the source entering a bright Hard state or 
counter-clockwise motion in the hardness-intensity diagram.

\section{Connections between Hysteresis, Jets, and $R_{\rm in}$}

It is clear that physical processes are operating in BHC systems
that are not captured by canonical state definitions and the standard
physical picture that changes in emission properties are driven only
by changes in mass accretion rate.  Observations of hysteresis of 
state transitions and radio jets are examples of phenomena that 
do not have good physical explanations.  Observations of
4U~1630--47 as well as other BHC systems suggest that the two 
phenomena may be related.  For GX~339--4, XTE J1859+226, 
XTE J1550--564, and 4U~1630--47 in 1998, major radio ejections
are preceded by bright Hard states, while the recent outburst
from 4U~1630--47 showed no bright Hard state (i.e., hysteresis)
and no major radio ejection.

Additionally, recent theoretical work suggests that both
phenomena (hysteresis and major radio ejections) may be regulated
by the location of the inner radius of the optically thick 
accretion disk, $R_{\rm in}$.  \cite{mlm05} suggest that if
$R_{\rm in}$ becomes very large when sources enter quiescence, 
allowing for a central corona with strong hard X-ray emission, 
then this hard X-ray emission may inhibit the formation of the
inner optically thick disk.  If this occurs, then the hard 
X-ray source (and thus the Hard state) may survive up to 
very high accretion rates.  The ``unified model'' for BHC
jets by \cite{fbg04} has been inspired, in part, by the
Transition-based state definitions.  In this model, the
main physical parameter that determines the nature of the
outflow is also $R_{\rm in}$.  The slow, steady compact 
jets are produced when $R_{\rm in}$ is large, while the
disk must approach the black hole (small $R_{\rm in}$)
before the highly relativistic major outflows can be
produced.  Although more work is required to show that
$R_{\rm in}$ is indeed the other important physical 
parameter in setting the BHC emission properties, understanding
the accretion geometries and how they are related to 
the different kinds of jets would be a major step forward 
in our understanding of the disk/jet connection.

\vspace{1.0cm}
This paper is based on a talk given at the COSPAR colloquium 
``Spectra and Timing of Compact X-Ray Binaries'' in Mumbai, 
India.  I would like to thank the organizers, especially 
Prof. Pranab Ghosh, for inviting me to speak and for their
hospitality at the meeting.  I would like to thank Tomaso 
Belloni, Stephane Corbel, Philip Kaaret, Andrea Goldwurm,
Joern Wilms, Rick Rothschild, and Katja Pottschmidt for 
discussions that have been useful in the preparation of
this work.  I acknowledge partial support from NASA grants 
NAG5-13055, NNG04GA49G, and NAG5-12703.




\begin{thebibliography}{41}
\expandafter\ifx\csname natexlab\endcsname\relax\def\natexlab#1{#1}\fi
\expandafter\ifx\csname url\endcsname\relax
  \def\url#1{\texttt{#1}}\fi
\expandafter\ifx\csname urlprefix\endcsname\relax\def\urlprefix{URL }\fi

\bibitem[{{Abramowicz} et~al.(1988){Abramowicz}, {Czerny}, {Lasota}, and
  {Szuszkiewicz}}]{abramowicz88}
{Abramowicz}, M.~A., {Czerny}, B., {Lasota}, J.~P., {Szuszkiewicz}, E.,
  September 1988. {Slim Accretion Disks}. ApJ 332, 646--658.

\bibitem[{{Belloni}(2005)}]{belloni05}
{Belloni}, T., April 2005. {Black Hole States: Accretion and Jet Ejection}.
  astro-ph/0504185.

\bibitem[{{Belloni} et~al.(1996){Belloni}, {Mendez}, {van der Klis},
  {Hasinger}, {Lewin}, and {van Paradijs}}]{belloni96}
{Belloni}, T., {Mendez}, M., {van der Klis}, M., {Hasinger}, G., {Lewin},
  W.~H.~G., {van Paradijs}, J., December 1996. {An Intermediate State of Cygnus
  X-1}. ApJ 472, L107.

\bibitem[{{Blandford} and {Begelman}(1999)}]{bb99}
{Blandford}, R.~D., {Begelman}, M.~C., February 1999. {On the Fate of Gas
  Accreting at a Low Rate on to a Black Hole}. MNRAS 303, L1--L5.

\bibitem[{{Chakrabarti} and {Titarchuk}(1995)}]{ct95}
{Chakrabarti}, S., {Titarchuk}, L.~G., December 1995. {Spectral Properties of
  Accretion Disks around Galactic and Extragalactic Black Holes}. ApJ 455, 623.

\bibitem[{{Coppi}(1992)}]{coppi92}
{Coppi}, P.~S., October 1992. {Time-dependent models of magnetized pair
  plasmas}. MNRAS 258, 657--683.

\bibitem[{{Corbel} et~al.(2004){Corbel}, {Fender}, {Tomsick}, {Tzioumis}, and
  {Tingay}}]{corbel04}
{Corbel}, S., {Fender}, R.~P., {Tomsick}, J.~A., {Tzioumis}, A.~K., {Tingay},
  S., December 2004. {On the Origin of Radio Emission in the X-Ray States of
  XTE J1650-500 during the 2001-2002 Outburst}. ApJ 617, 1272--1283.

\bibitem[{{Corbel} et~al.(2000){Corbel}, {Fender}, {Tzioumis}, {Nowak},
  {McIntyre}, {Durouchoux}, and {Sood}}]{corbel00}
{Corbel}, S., {Fender}, R.~P., {Tzioumis}, A.~K., {Nowak}, M., {McIntyre}, V.,
  {Durouchoux}, P., {Sood}, R., July 2000. {Coupling of the X-ray and Radio
  Emission in the Black Hole Candidate and Compact Jet Source GX 339--4}. A\&A
  359, 251--268.

\bibitem[{{Dhawan} et~al.(2000){Dhawan}, {Mirabel}, and {Rodr{\'
  i}guez}}]{dhawan00}
{Dhawan}, V., {Mirabel}, I.~F., {Rodr{\' i}guez}, L.~F., November 2000.
  {AU-Scale Synchrotron Jets and Superluminal Ejecta in GRS 1915+105}. ApJ 543,
  373--385.

\bibitem[{{Dieters} et~al.(2000){Dieters}, {Belloni}, {Kuulkers}, {Woods},
  {Cui}, {Zhang}, {Chen}, {van der Klis}, {van Paradijs}, {Swank}, {Lewin}, and
  {Kouveliotou}}]{dieters00}
{Dieters}, S.~W., {Belloni}, T., {Kuulkers}, E., {Woods}, P., {Cui}, W.,
  {Zhang}, S.~N., {Chen}, W., {van der Klis}, M., {van Paradijs}, J., {Swank},
  J., {Lewin}, W.~H.~G., {Kouveliotou}, C., July 2000. {The Timing Evolution of
  4U 1630--47 during Its 1998 Outburst}. ApJ 538, 307--314.

\bibitem[{{Esin} et~al.(1997){Esin}, {McClintock}, and {Narayan}}]{emn97}
{Esin}, A.~A., {McClintock}, J.~E., {Narayan}, R., November 1997.
  {Advection-dominated Accretion and the Spectral States of Black Hole X-Ray
  Binaries: Application to Nova Muscae 1991}. ApJ 489, 865.

\bibitem[{{Fender}(2001)}]{fender01}
{Fender}, R.~P., March 2001. {Powerful Jets from Black Hole X-Ray Binaries in
  Low/Hard X-Ray States}. MNRAS 322, 31--42.

\bibitem[{{Fender} et~al.(2004){Fender}, {Belloni}, and {Gallo}}]{fbg04}
{Fender}, R.~P., {Belloni}, T.~M., {Gallo}, E., October 2004. {Towards a
  Unified Model for Black Hole X-Ray Binary Jets}. MNRAS, 538.

\bibitem[{{Fender} et~al.(2003){Fender}, {Gallo}, and {Jonker}}]{fgj03}
{Fender}, R.~P., {Gallo}, E., {Jonker}, P.~G., August 2003. {Jet-Dominated
  States: An Alternative to Advection across Black Hole Event Horizons in
  `Quiescent' X-Ray Binaries}. MNRAS 343, L99--L103.

\bibitem[{{Gallo} et~al.(2004){Gallo}, {Corbel}, {Fender}, {Maccarone}, and
  {Tzioumis}}]{gallo04}
{Gallo}, E., {Corbel}, S., {Fender}, R.~P., {Maccarone}, T.~J., {Tzioumis},
  A.~K., January 2004. {A Transient Large-Scale Relativistic Radio Jet from 
  GX 339--4}. MNRAS 347, L52--L56.

\bibitem[{{Gallo} et~al.(2003){Gallo}, {Fender}, and {Pooley}}]{gfp03}
{Gallo}, E., {Fender}, R.~P., {Pooley}, G.~G., September 2003. {A Universal
  Radio-X-Ray Correlation in Low/Hard State Black Hole Binaries}. MNRAS 344,
  60--72.

\bibitem[{{Grove} et~al.(1998){Grove}, {Johnson}, {Kroeger}, {McNaron-Brown},
  {Skibo}, and {Phlips}}]{grove98}
{Grove}, J.~E., {Johnson}, W.~N., {Kroeger}, R.~A., {McNaron-Brown}, K.,
  {Skibo}, J.~G., {Phlips}, B.~F., June 1998. {Gamma-Ray Spectral States of
  Galactic Black Hole Candidates}. ApJ 500, 899.

\bibitem[{{Hannikainen} et~al.(2002){Hannikainen}, {Sault}, {Kuulkers}, {Wu},
  {Jones}, and {Hunstead}}]{hannikainen02}
{Hannikainen}, D., {Sault}, B., {Kuulkers}, E., {Wu}, K., {Jones}, P.,
  {Hunstead}, R., September 2002. {Follow-Up Radio Observations of the Black
  Hole Candidate 4U 1630--47}. The Astronomer's Telegram 108.

\bibitem[{{Hjellming} et~al.(1999){Hjellming}, {Rupen}, {Mioduszewski},
  {Kuulkers}, {McCollough}, {Harmon}, {Buxton}, {Sood}, {Tzioumis}, {Rayner},
  {Dieters}, and {Durouchoux}}]{hjellming99}
{Hjellming}, R.~M., {Rupen}, M.~P., {Mioduszewski}, A.~J., {Kuulkers}, E.,
  {McCollough}, M., {Harmon}, B.~A., {Buxton}, M., {Sood}, R., {Tzioumis}, A.,
  {Rayner}, D., {Dieters}, S., {Durouchoux}, P., March 1999. {Radio and X-Ray
  Observations of the 1998 Outburst of the Recurrent X-Ray Transient 4U
  1630--47}. ApJ 514, 383--387.

\bibitem[{{Hjellming} and {Wade}(1971)}]{hw71}
{Hjellming}, R.~M., {Wade}, C.~M., August 1971. {Radio Emission from X-Ray
  Sources}. ApJ 168, L21.

\bibitem[{{Homan} and {Belloni}(2004)}]{hb04}
{Homan}, J., {Belloni}, T., December 2004. {The Evolution of Black Hole
  States}. astro-ph/0412597.

\bibitem[{{Homan} et~al.(2001){Homan}, {Wijnands}, {van der Klis}, {Belloni},
  {van Paradijs}, {Klein-Wolt}, {Fender}, and {M{\'e}ndez}}]{homan00}
{Homan}, J., {Wijnands}, R., {van der Klis}, M., {Belloni}, T., {van Paradijs},
  J., {Klein-Wolt}, M., {Fender}, R., {M{\'e}ndez}, M., February 2001.
  {Correlated X-Ray Spectral and Timing Behavior of the Black Hole Candidate
  XTE J1550--564: A New Interpretation of Black Hole States}. ApJS 132,
  377--402.

\bibitem[{{Markoff} and {Nowak}(2004)}]{mn04}
{Markoff}, S., {Nowak}, M.~A., July 2004. {Constraining X-Ray Binary Jet Models
  via Reflection}. ApJ 609, 972--976.

\bibitem[{McClintock and Remillard(2003)}]{mr03}
McClintock, J., Remillard, R., 2003. {Black Hole Binaries}. Review Article
  astro-ph/0306213.

\bibitem[{{Mendez} and {van der Klis}(1997)}]{mendez97}
{Mendez}, M., {van der Klis}, M., April 1997. {The EXOSAT Data on GX 339--4:
  Further Evidence for an ``Intermediate'' State}. ApJ 479, 926.

\bibitem[{{Meyer-Hofmeister} et~al.(2005){Meyer-Hofmeister}, {Liu}, and
  {Meyer}}]{mlm05}
{Meyer-Hofmeister}, E., {Liu}, B.~F., {Meyer}, F., March 2005. {Hysteresis in
  Spectral State Transitions - A Challenge for Theoretical Modeling}. A\&A 432,
  181--187.

\bibitem[{{Mirabel} and {Rodr{\' i}guez}(1999)}]{mr99}
{Mirabel}, I.~F., {Rodr{\' i}guez}, L.~F., 1999. {Sources of Relativistic Jets
  in the Galaxy}. ARA\&A 37, 409--443.

\bibitem[{{Mitsuda} et~al.(1984){Mitsuda}, {Inoue}, {Koyama}, {Makishima},
  {Matsuoka}, {Ogawara}, {Suzuki}, {Tanaka}, {Shibazaki}, and
  {Hirano}}]{mitsuda84}
{Mitsuda}, K., {Inoue}, H., {Koyama}, K., {Makishima}, K., {Matsuoka}, M.,
  {Ogawara}, Y., {Suzuki}, K., {Tanaka}, Y., {Shibazaki}, N., {Hirano}, T.,
  1984. {Energy Spectra of Low-Mass Binary X-Ray Sources Observed from TENMA}.
  PASJ 36, 741--759.

\bibitem[{{Miyamoto} et~al.(1991){Miyamoto}, {Kimura}, {Kitamoto}, {Dotani},
  and {Ebisawa}}]{miyamoto91}
{Miyamoto}, S., {Kimura}, K., {Kitamoto}, S., {Dotani}, T., {Ebisawa}, K.,
  December 1991. {X-Ray Variability of GX 339--4 in its Very High State}. ApJ
  383, 784--807.

\bibitem[{{Shakura} and {Sunyaev}(1973)}]{ss73}
{Shakura}, N.~I., {Sunyaev}, R.~A., 1973. {Black Holes in Binary Systems.
  Observational Appearance}. A\&A 24, 337--355.

\bibitem[{{Stirling} et~al.(2001){Stirling}, {Spencer}, {de la Force},
  {Garrett}, {Fender}, and {Ogley}}]{stirling01}
{Stirling}, A.~M., {Spencer}, R.~E., {de la Force}, C.~J., {Garrett}, M.~A.,
  {Fender}, R.~P., {Ogley}, R.~N., November 2001. {A Relativistic Jet from
  Cygnus X-1 in the Low/Hard X-Ray State}. MNRAS 327, 1273--1278.

\bibitem[{{Tanaka} and {Shibazaki}(1996)}]{ts96}
{Tanaka}, Y., {Shibazaki}, N., 1996. {X-Ray Novae}. ARA\&A 34, 607--644.

\bibitem[{{Tananbaum} et~al.(1972){Tananbaum}, {Gursky}, {Kellogg}, {Giacconi},
  and {Jones}}]{tananbaum72}
{Tananbaum}, H., {Gursky}, H., {Kellogg}, E., {Giacconi}, R., {Jones}, C.,
  October 1972. {Observation of a Correlated X-Ray Transition in Cygnus X-1}.
  ApJ 177, L5.

\bibitem[{{Tomsick}(2004)}]{tomsick04_rossi}
{Tomsick}, J.~A., July 2004. {The Evolution of Accreting Black Holes in
  Outburst}. In: AIP Conf. Proc. 714: X-ray Timing 2003: Rossi and Beyond,
  astro-ph/0401189. pp. 71--78.

\bibitem[{{Tomsick} et~al.(2005){Tomsick}, {Corbel}, {Goldwurm}, and
  {Kaaret}}]{tomsick05}
{Tomsick}, J.~A., {Corbel}, S., {Goldwurm}, A., {Kaaret}, P., May 2005. {X-Ray
  Observations of the Black Hole Transient 4U 1630--47 During Two Years of X-Ray
  Activity}. astro-ph/0505271, Accepted by ApJ.

\bibitem[{{Tomsick} and {Kaaret}(2000)}]{tk00}
{Tomsick}, J.~A., {Kaaret}, P., July 2000. {X-Ray Spectral and Timing Evolution
  during the Decay of the 1998 Outburst from the Recurrent X-Ray Transient 4U
  1630--47}. ApJ 537, 448--460.

\bibitem[{{Tomsick} et~al.(1999){Tomsick}, {Kaaret}, {Kroeger}, and
  {Remillard}}]{tomsick99}
{Tomsick}, J.~A., {Kaaret}, P., {Kroeger}, R.~A., {Remillard}, R.~A., February
  1999. {Broadband X-Ray Spectra of the Black Hole Candidate GRO J1655--40}.
  ApJ 512, 892--900.

\bibitem[{{Trudolyubov} et~al.(2001){Trudolyubov}, {Borozdin}, and
  {Priedhorsky}}]{tbp01}
{Trudolyubov}, S.~P., {Borozdin}, K.~N., {Priedhorsky}, W.~C., April 2001.
  {{\em RXTE} Observations of 4U 1630--47 during the peak of its 1998 outburst}.
  MNRAS 322, 309--320.

\bibitem[{{van der Klis}(2004)}]{vdk04}
{van der Klis}, M., October 2004. {A Review of Rapid X-Ray Variability in X-Ray
  Binaries}. astro-ph/0410551.

\bibitem[{{van der Klis}(2004)}]{vdk95}
{van der Klis}, M., 1995. {Rapid Aperiodic Variability in X-Ray Binaries}. 
In: X-ray Binaries. Cambridge University Press. pp.252-307.

\bibitem[{{Wilson} and {Rothschild}(1983)}]{wr83}
{Wilson}, C.~K., {Rothschild}, R.~E., November 1983. {Observations of a Hard
  X-Ray Component in the Spectrum of Nova Ophiuchi}. ApJ 274, 717--722.

\end{thebibliography}

\end{document}